# First observation of $\Lambda_c^+ \to \Lambda K^+\pi^0$ and evidence of $\Lambda_c^+ \to \Lambda K^+\pi^+\pi^-$

M. Ablikim et al.*
(BESIII Collaboration)

We present the first observation of the singly Cabibbo-suppressed decay $\Lambda_c^+ \to \Lambda K^+\pi^0$ with a significance of $5.7\sigma$ and the first evidence of $\Lambda_c^+ \to \Lambda K^+\pi^+\pi^-$ decay with a significance of $3.1\sigma$, based on $e^+e^-$ annihilation data recorded by the BESIII detector at the BEPCII collider. The data correspond to an integrated luminosity of 6.4 $\mathrm{fb}^{-1}$, in the center-of-mass energy range from 4.600 to 4.950 GeV. We determine the branching fractions of $\Lambda_c^+ \to \Lambda K^+\pi^0$ and $\Lambda_c^+ \to \Lambda K^+\pi^+\pi^-$ relative to their Cabibbo-favored counterparts to be $\frac{\mathcal{B}(\Lambda_c^+\to\Lambda K^+\pi^0)}{\mathcal{B}(\Lambda_c^+\to\Lambda\pi^+\pi^0)} = (2.09 \pm 0.39_{\mathrm{stat}} \pm 0.07_{\mathrm{syst}}) \times 10^{-2}$ and $\frac{\mathcal{B}(\Lambda_c^+\to\Lambda K^+\pi^+\pi^-)}{\mathcal{B}(\Lambda_c^+\to\Lambda\pi^+\pi^+\pi^-)} = (1.13 \pm 0.41_{\mathrm{stat}} \pm 0.06_{\mathrm{syst}}) \times 10^{-2}$, respectively. Moreover, by combining our measured result with the world average of $\mathcal{B}(\Lambda_c^+ \to \Lambda\pi^+\pi^0)$, we obtain the branching fraction $\mathcal{B}(\Lambda_c^+ \to \Lambda K^+\pi^0) = (1.49 \pm 0.27_{\mathrm{stat}} \pm 0.05_{\mathrm{syst}} \pm 0.08_{\mathrm{ref}}) \times 10^{-3}$. This result significantly departs from theoretical predictions based on quark $SU(3)$ flavor symmetry, which is underpinned by the presumption of meson pair $S$-wave amplitude dominance.

## I. INTRODUCTION

Since its discovery in the 1970s [1,2], the $\Lambda_c^+$ charmed baryon has remained a focal point of particle-physics research. The $\Lambda_c^+$ presents a unique opportunity to probe the dynamics of light quarks within a three-body structure, coexisting with a heavy $c$ quark. Unlike charmed-meson decays, where nonfactorizable effects are typically negligible, the decays of $\Lambda_c^+$ involve a more substantial contribution from internal $W$-emission and $W$-exchange diagrams, as depicted in Fig. 1. Consequently, the decay mechanisms of $\Lambda_c^+$ are more intricate [3,4], and experimental studies of $\Lambda_c^+$ decays are of critical importance in testing various theoretical models and illuminating the underlying dynamics.

Despite their significance, singly Cabibbo-suppressed (SCS) decays of $\Lambda_c^+$ have been less explored than Cabibbo-favored (CF) decays, primarily due to their low branching fractions (BFs) and the limited data samples available. Recently, BESIII studied the two-body SCS decays $\Lambda_c^+ \to n\pi^+$ [5] and $\Lambda_c^+ \to \Lambda K^+$ [6]. The measured BF of $\Lambda_c^+ \to n\pi^+$ is twice larger than the predicted value [7], suggesting that the nonfactorizable contribution in this decay is overestimated. However, the measured BF of $\Lambda_c^+ \to \Lambda K^+$ is around 40% of the expectations derived from quark $SU(3)$ flavor symmetry [8], constituent quark models [9] or current algebra [7], indicating that the nonfactorizable contribution in this decay is poorly estimated. Experimental investigations of additional SCS decays of $\Lambda_c^+$ are essential for improving our understanding of the mechanisms responsible for the behavior of this baryon.

To date, the decay $\Lambda_c^+ \to \Lambda K^+\pi^0$ has not been observed in any experiment. Based on quark $SU(3)$ flavor symmetry, the $S$-wave meson pair $MM'$ configurations, where $M(M')$ denotes the meson octets, are assumed to dominate in the final state, and the BF of this decay is predicted to be $(4.5 \pm 0.8) \times 10^{-3}$ in Ref. [10] and $(3.5 \pm 0.6) \times 10^{-3}$ in Ref. [11], where the latter result incorporates an additional constraint stemming from the magnitude of the $S$-wave and $P$-wave interference term $\alpha$ [12]. In Ref. [13] the *BABAR* experiment report a null search for $\Lambda_c^+ \to \Lambda K^+\pi^+\pi^-$, in which an upper limit on the BF ratio $\frac{\mathcal{B}(\Lambda_c^+\to\Lambda K^+\pi^+\pi^-)}{\mathcal{B}(\Lambda_c^+\to\Lambda\pi^+)}$ is set to be $4.1 \times 10^{-2}$ at the 90% confidence level.

In this paper, we report the first observation of $\Lambda_c^+ \to \Lambda K^+\pi^0$ and evidence of the decay $\Lambda_c^+ \to \Lambda K^+\pi^+\pi^-$, using $e^+e^-$ collision data, at 13 c.m. energies ranging from 4.600 to 4.950 GeV [14,15] corresponding to an integral luminosity of 6.4 $\mathrm{fb}^{-1}$, collected with the BESIII detector at the BEPCII collider. Their BFs are measured by normalizing to those of the CF decays $\Lambda_c^+ \to \Lambda\pi^+\pi^0$ and $\Lambda_c^+ \to \Lambda\pi^+\pi^+\pi^-$, respectively. Charge conjugation is always implied throughout this paper unless explicitly mentioned.

*Full author list given at the end of the article.

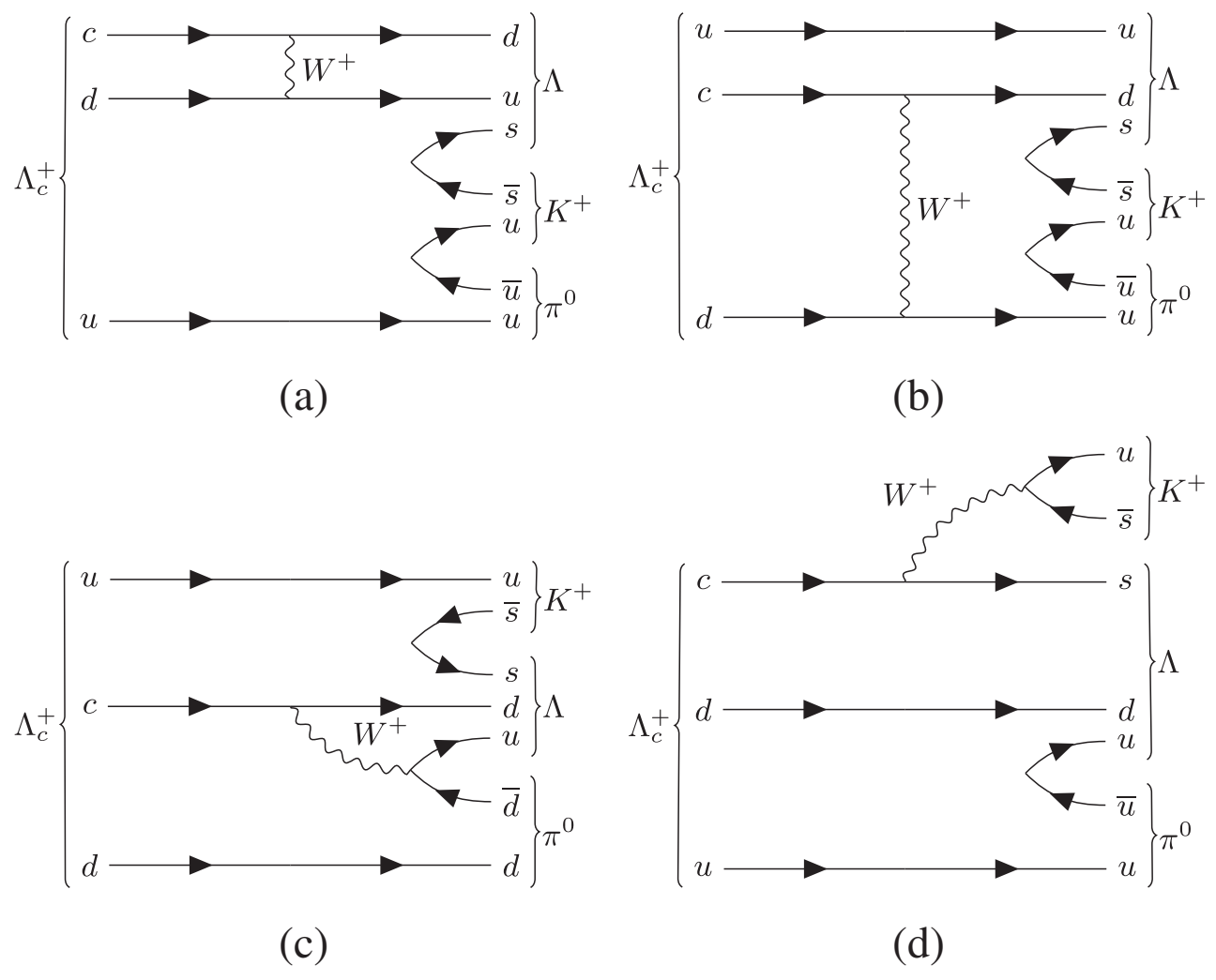

FIG. 1. Typical Feynman diagrams for $\Lambda_c^+ \to \Lambda K^+ \pi^0$: (a) and (b) $W$-exchange diagrams, (c) internal $W$-emission diagram, and (d) external $W$-emission diagram.

## II. DETECTOR AND SIMULATION

The BESIII detector [16] records symmetric $e^+e^-$ collisions provided by the BEPCII storage ring [17] in the c.m. energy range from 2.0 to 4.95 GeV, with a peak luminosity of $1 \times 10^{33}$ cm$^{-2}$ s$^{-1}$ achieved at $\sqrt{s} = 3.77$ GeV. BESIII has collected large data samples in these energy regions [18–22]. The cylindrical core of the BESIII detector covers 93% of the full solid angle and consists of a helium-based multilayer drift chamber (MDC), a plastic scintillator time-of-flight system (TOF), and a CsI(Tl) electromagnetic calorimeter (EMC), which are all enclosed in a superconducting solenoidal magnet providing a 1.0 T magnetic field. The solenoid is supported by an octagonal flux-return yoke with resistive plate counter muon identification modules interleaved with steel. The charged-particle momentum resolution at 1 GeV/$c$ is 0.5%, and the d$E$/d$x$ resolution is 6% for electrons from Bhabha scattering. The EMC measures photon energies with a resolution of 2.5% (5%) at 1 GeV in the barrel (end cap) region. The time resolution in the TOF barrel region is 68 ps, while that in the end cap region is 60 ps [23].

Monte Carlo (MC) simulation is performed with GEANT4 based software [24], which contains a description of the geometry and response of the BESIII detector [25]. To estimate detection efficiency, the KKMC generator [26] is used to generate signal MC samples. This generator includes the effects of initial-state radiation and the beam-energy spread, and incorporates the Born cross section line shape of $e^+e^- \to \Lambda_c^+ \bar{\Lambda}_c^-$ measured by BESIII [27]. In these signal MC samples, the $\Lambda_c^+$ is set to decay via the exclusive modes, $\Lambda_c^+ \to \Lambda K^+ \pi^0$, $\Lambda_c^+ \to \Lambda K^+ \pi^+ \pi^-$, $\Lambda_c^+ \to \Lambda \pi^+ \pi^0$, and $\Lambda_c^+ \to \Lambda \pi^+ \pi^+ \pi^-$, while the $\bar{\Lambda}_c^-$ decays inclusively according to BFs taken from the Particle Data Group (PDG) [28]. The simulation samples for the $\Lambda_c^+ \to \Lambda K^+ \pi^0$ and $\Lambda_c^+ \to \Lambda K^+ \pi^+ \pi^-$ decays are produced with a phase-space model. The resonance structures in the decays $\Lambda_c^+ \to \Lambda \pi^+ \pi^0$ and $\Lambda_c^+ \to \Lambda \pi^+ \pi^+ \pi^-$ are modeled according to the observed decay patterns in data. Exclusive MC samples of $\Lambda_c^+ \to \Xi^0 K^+$, $\Lambda_c^+ \to \Lambda K^+ K_S^0$, $\Lambda_c^+ \to \Xi^- K^+ \pi^+$, and $\Lambda_c^+ \to \Xi(1530) K^+$ decays are generated for background studies, together with an inclusive MC sample, consisting of $\Lambda_c^+ \bar{\Lambda}_c^-$, QED related and hadron production [29]. The subsequent decays of all the intermediate states in the MC samples are simulated by EvtGen [30], using BFs either taken from the PDG [28], when available, or otherwise estimated with LUNDCHARM [31,32]. Final-state radiation from charged final state particles is incorporated using PHOTOS [33].

## III. EVENT SELECTION

The majority of the dataset used in the analysis is situated near the $\Lambda_c^+ \bar{\Lambda}_c^-$ threshold, where the production of $\Lambda_c^+ \bar{\Lambda}_c^-$ pairs without associated hadrons is prevalent. This environment lends itself to the adoption of the single-tag method, where only one $\Lambda_c^+$ is reconstructed within an event, with no condition on the recoil side. This approach is favored for its efficiency, thereby enabling the retention of a greater number of $\Lambda_c^+$ candidates.

Charged tracks detected in the MDC are required to be within a polar angle ($\theta$) range of $|\cos\theta| < 0.93$, where $\theta$ is defined with respect to the $z$ axis, which is the symmetry axis of the MDC. For charged tracks not originating from $\Lambda$ decays, the nearest distance between tracks to the $e^+e^-$ interaction point (IP) must be no more than 10 cm along the $z$ axis, $|V_z|$, and less than 1 cm in the transverse plane, $|V_{\rm xy}|$. Particle identification (PID) is implemented by combining information on the specific ionization energy loss in the MDC (d$E$/d$x$) and the flight time in the TOF to form the likelihoods $\mathcal{L}(h)$ ($h = p, \pi, K$) for each hadron $h$ hypothesis. Tracks are identified as protons when the proton hypothesis satisfies the requirements $\mathcal{L}(p) > \mathcal{L}(\pi)$ and $\mathcal{L}(p) > \mathcal{L}(K)$. Charged pions and kaons are discriminated based on comparing the likelihoods for the hypotheses, $\mathcal{L}(\pi) > \mathcal{L}(K)$ and $\mathcal{L}(K) > \mathcal{L}(\pi)$, respectively. The $\Lambda$ candidates are reconstructed from a pair of oppositely charged proton and pion candidates, identified with relatively loose PID requirements. In this case, the charged tracks must have a closest distance to the IP within $\pm 20$ cm, with no transverse distance requirement imposed. The daughter tracks are constrained to originate from the same decay vertex with a $\chi^2$ value less than 100, and this vertex is required to be displaced from the IP by a distance at least twice larger than the measurement uncertainty. It is demanded that the $\Lambda$ candidates have a $p\pi^-$ invariant mass within $1.111 < M_{p\pi^-} < 1.121$ GeV/$c^2$, which corresponds to three standard deviations of the reconstruction resolution around the known $\Lambda$ mass [28].

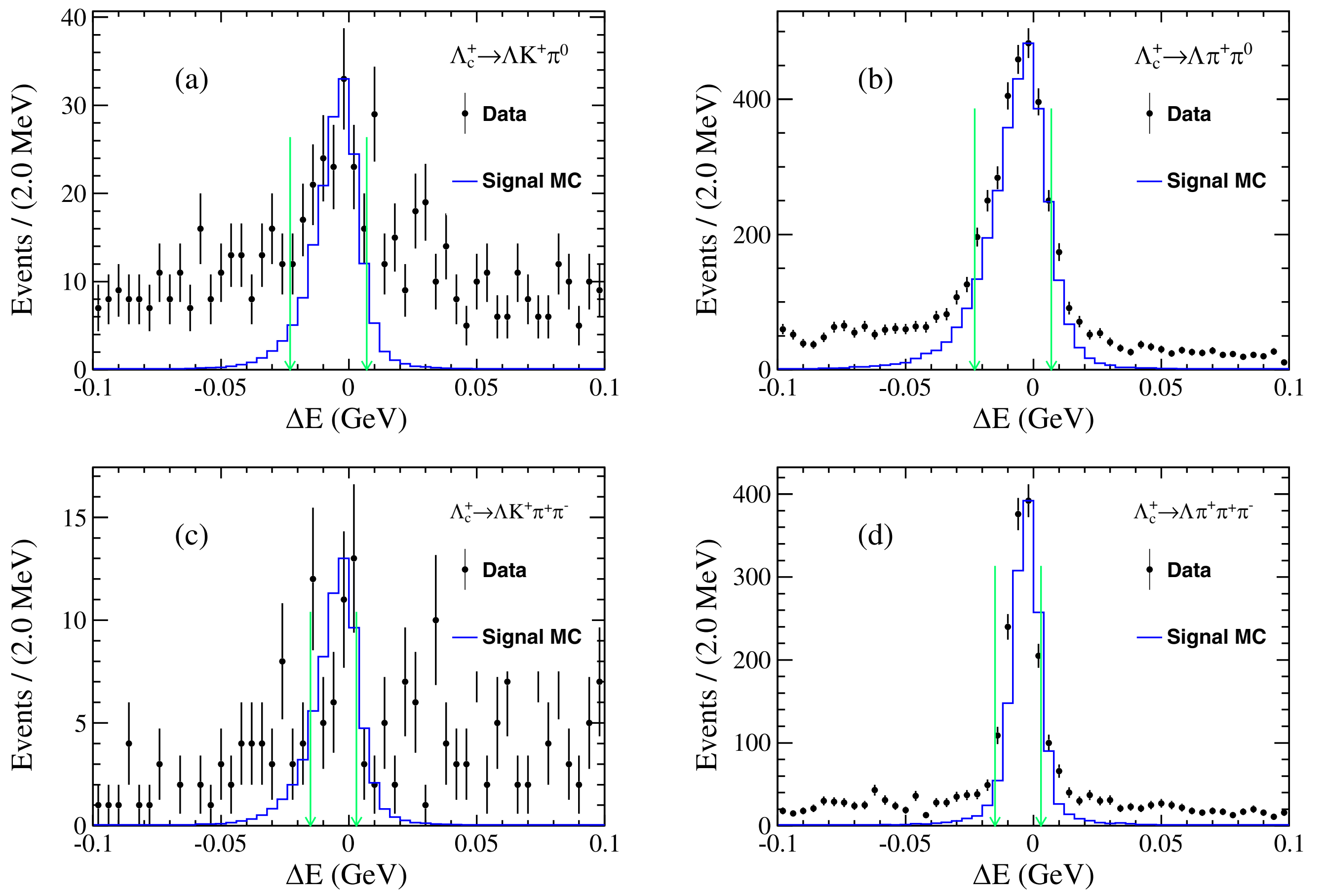

FIG. 2. Distributions of $\Delta E$ for (a) $\Lambda_c^+ \to \Lambda K^+\pi^0$, (b) $\Lambda_c^+ \to \Lambda\pi^+\pi^0$, (c) $\Lambda_c^+ \to \Lambda K^+\pi^+\pi^-$, and (d) $\Lambda_c^+ \to \Lambda\pi^+\pi^+\pi^-$. The teal arrows are the optimized $\Delta E$ windows. All events have been imposed with all other selection criteria and an additional requirement of $M_{\rm BC} \in [2.282, 2.291]$ GeV$/c^2$. The signal MC histograms are normalized so that the heights of the peaks agree with those in data.

Electromagnetic showers produced in the EMC, not associated with any charged tracks, are identified as photon candidates. The deposited energy is required to be greater than 25 MeV in the barrel region ($|\cos\theta| < 0.80$) and greater than 50 MeV in the end cap region ($0.86 < |\cos\theta| < 0.92$). To further suppress background from beam and electronic noise, the difference of EMC time with respect to the collision time is required to be within 700 ns. Showers are required to be separated from other charged tracks by an angle greater than 10° in order to eliminate activity induced by tracks. Then, the $\pi^0$ candidates are reconstructed from photon pairs with invariant mass within $0.115 < M_{\gamma\gamma} < 0.150$ GeV$/c^2$. To improve the $\pi^0$ momentum resolution, the mass of the $\pi^0$ candidate is constrained to the PDG value [28] via a one-constraint kinematic fit. Combinations satisfying $\chi^2 < 200$ are preserved, and the refined momenta are utilized for subsequent studies.

By analyzing the inclusive MC samples with the tool TOPOANA [34], we find several processes with the same final states as the signals contaminate the selection. For $\Lambda_c^+ \to \Lambda K^+\pi^0$, we veto events where the invariant mass of the $\Lambda\pi^0$ pair satisfies $1.290 < M_{\Lambda\pi^0} < 1.340$ GeV$/c^2$ to suppress background from $\Lambda_c^+ \to \Xi^0 K^+$ decays. For $\Lambda_c^+ \to \Lambda K^+\pi^+\pi^-$, we reject candidates with $1.310 < M_{\Lambda\pi^-} < 1.330$ GeV$/c^2$ and $0.490 < M_{\pi^+\pi^-} < 0.505$ GeV$/c^2$ to suppress contamination from $\Lambda_c^+ \to \Xi^- K^+\pi^+$, $\Lambda_c^+ \to \Xi(1530)K^+$, and $\Lambda_c^+ \to \Lambda K^+ K_S^0$ decays. In the selection of $\Lambda_c^+ \to \Lambda\pi^+\pi^+\pi^-$, events with $0.480 < M_{\pi^+\pi^-} < 0.520$ GeV$/c^2$ are discarded to suppress the background from $\Lambda_c^+ \to p K_S^0\pi^+\pi^-$ decays.

To further mitigate the effects of combinatorial background, two kinematic variables are employed: $\Delta E$ and the beam-constrained mass, $M_{\rm BC}$. The variable $\Delta E$ is defined as $E_{{\rm rec}-\Lambda_c^+} - E_{\rm beam}$, where $E_{{\rm rec}-\Lambda_c^+}$ represents the energy of the reconstructed $\Lambda_c^+$ and $E_{\rm beam}$ is the beam energy. The beam-constrained mass, $M_{\rm BC}$, is a crucial parameter used for determining signal yields. It is defined as $M_{\rm BC} \equiv \sqrt{E_{\rm beam}^2/c^4 - |\vec{p}_{\Lambda_c^+}|^2/c^2}$, where $p_{\Lambda_c^+}$ denotes the momentum of the reconstructed $\Lambda_c^+$. The $\Delta E$ distributions in the data are illustrated in Fig. 2. In case where multiple combinations exist, the one with the minimum $|\Delta E|$ is selected. For $\Lambda_c^+ \to \Lambda K^+\pi^0$ and $\Lambda_c^+ \to \Lambda\pi^+\pi^0$, candidates are required to satisfy $\Delta E \in [-0.023, 0.007]$ GeV. Meanwhile, for $\Lambda_c^+ \to \Lambda K^+\pi^+\pi^-$ and $\Lambda_c^+ \to \Lambda\pi^+\pi^+\pi^-$, the requirement is $\Delta E \in [-0.015, 0.003]$ GeV. These requirements have been optimized by maximizing the figure of merit, $\mathrm{FOM} = S/\sqrt{S+B}$, where $S$ is the expected signal yield in the signal region $M_{\mathrm{BC}} \in [2.282, 2.291]$ GeV/$c^2$ and $B$ is the background yield in the same region estimated from the inclusive MC sample. The number of $S(B)$ is normalized to the integrated luminosity of the data sample.

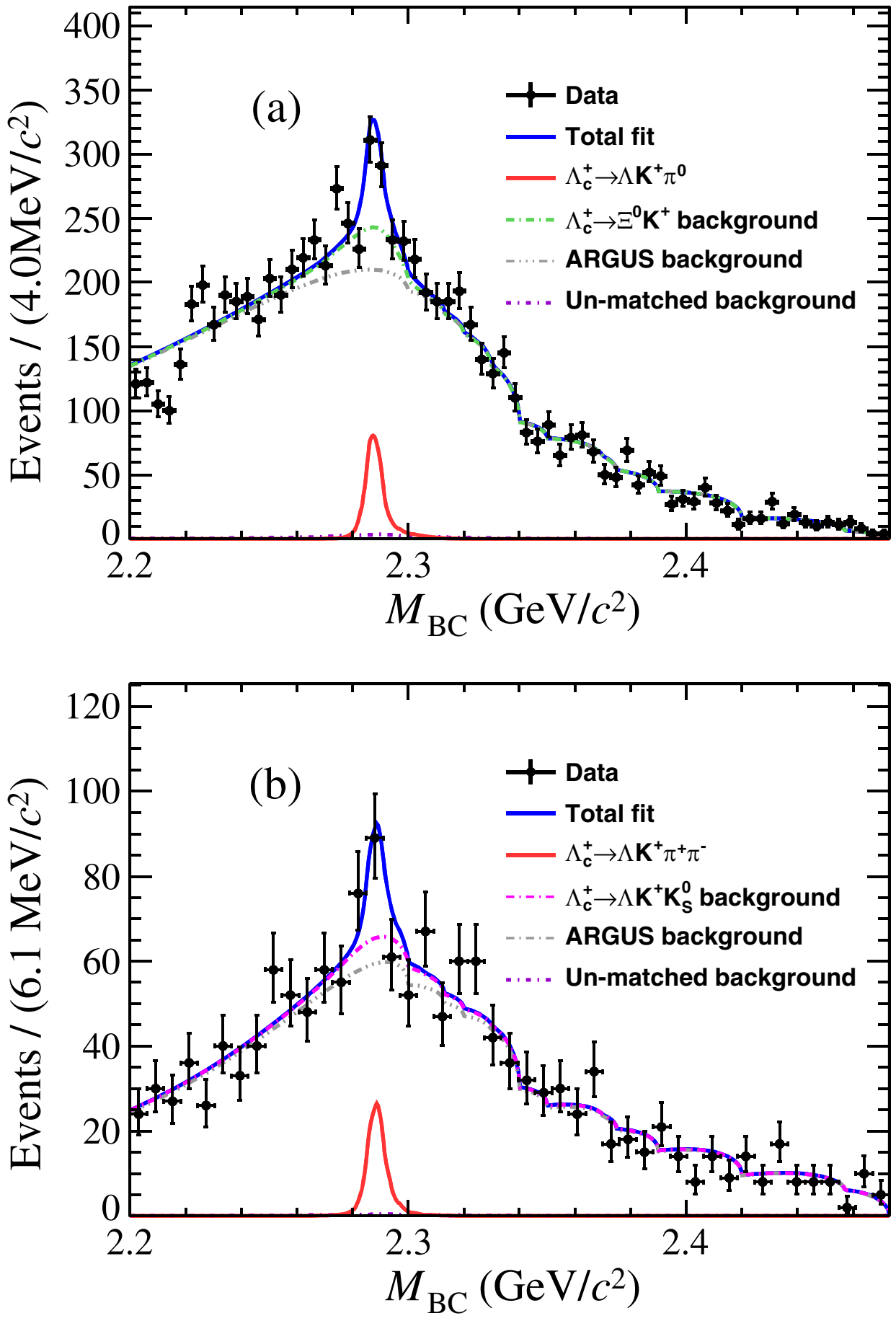

FIG. 3. Combined simultaneous fit results to the distributions of $M_{\mathrm{BC}}$ for (a) $\Lambda_c^+ \to \Lambda K^+\pi^0$ and (b) $\Lambda_c^+ \to \Lambda K^+\pi^+\pi^-$ at 13 energy points.

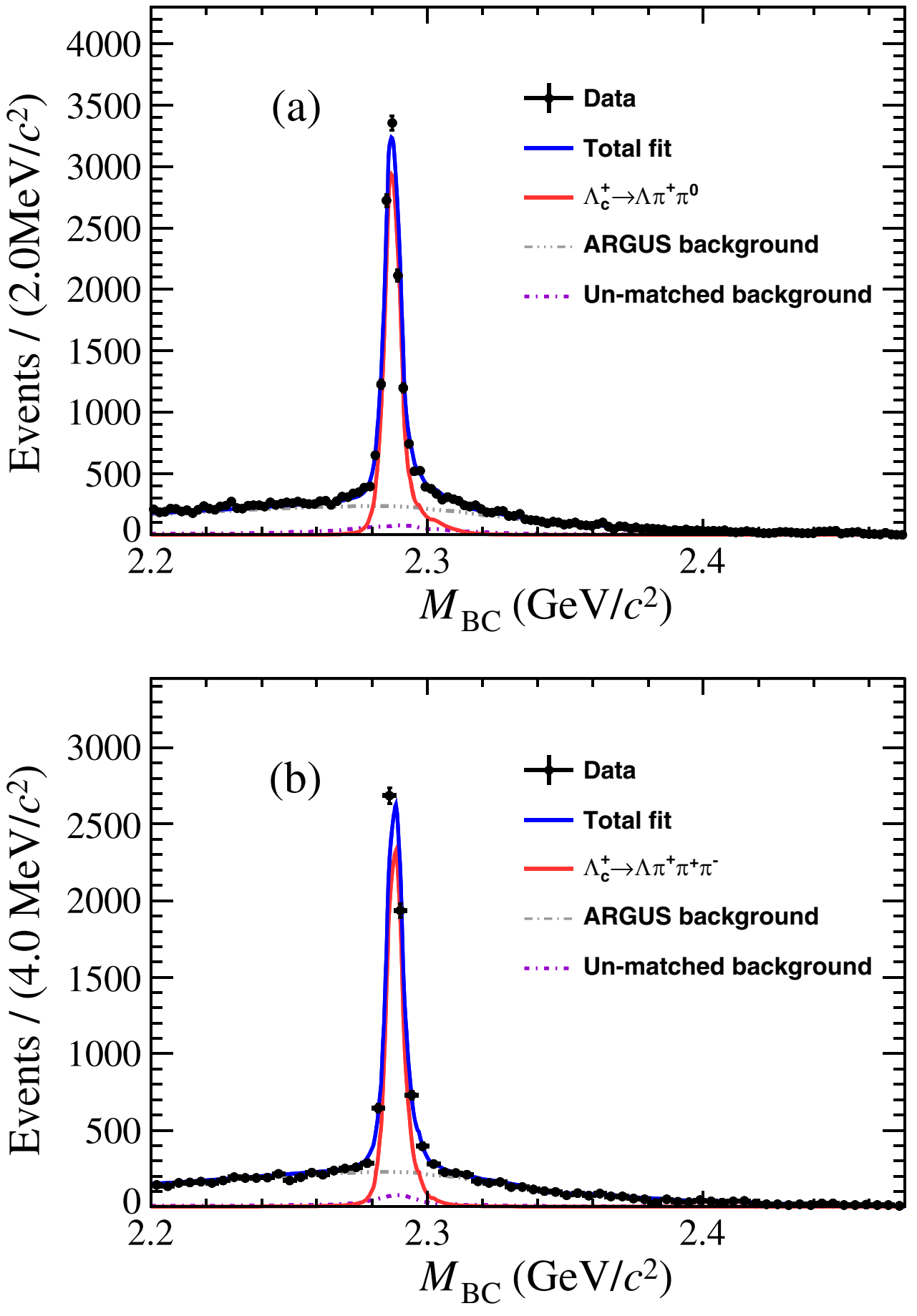

FIG. 4. Combined simultaneous fit results to the distributions of $M_{\mathrm{BC}}$ for (a) $\Lambda_c^+ \to \Lambda\pi^+\pi^0$ and (b) $\Lambda_c^+ \to \Lambda\pi^+\pi^+\pi^-$ at 13 energy points.

## IV. RELATIVE BF MEASUREMENT

Figure 3 shows the $M_{\mathrm{BC}}$ spectra of accepted events of $\Lambda_c^+ \to \Lambda K^+\pi^0$ and $\Lambda_c^+ \to \Lambda K^+\pi^+\pi^-$, and Fig. 4 shows the corresponding distribution for $\Lambda_c^+ \to \Lambda\pi^+\pi^0$ and $\Lambda_c^+ \to \Lambda\pi^+\pi^+\pi^-$ events. Signal peaks are evident for the SCS modes above the distributions of background events. To minimize systematic uncertainty, the BF of each signal decay is measured relative to its CF counterpart by

$$\mathcal{R} = \frac{\mathcal{B}^{\mathrm{sig}}}{\mathcal{B}^{\mathrm{ref}}} = \frac{N_i^{\mathrm{sig}} \cdot \epsilon_i^{\mathrm{ref}}}{N_i^{\mathrm{ref}} \cdot \epsilon_i^{\mathrm{sig}}}, \quad (1)$$

where $i$ represents each energy point, $N$ is the observed signal yield in data and $\epsilon$ denotes the detection efficiency obtained from signal MC samples. To determine $\mathcal{R}$, an unbinned maximum likelihood fit is performed on these $M_{\mathrm{BC}}$ spectra, in which $\mathcal{R}$ is a common fit parameter between the energy points. Table I lists the relative detection efficiencies and the signal yields of the reference modes.

To extract the signal yield of each mode, the simultaneous fit is performed on the $M_{\mathrm{BC}}$ distributions at different energy points. In the fit, the signal shapes of the four modes are described with the MC-simulated signal shapes convoluted with a Gaussian function that is used to compensate for the resolution difference between data and MC simulation. To obtain a pure signal, we employ the truth-match method. This method involves comparing the reconstructed tracks of two photons in the $\pi^0$ and the charged tracks $K^\pm$ and $\pi^\pm$ with their corresponding truth information.

TABLE I. Relative detection efficiencies for $\Lambda_c^+ \to \Lambda K^+\pi^0$ $(A)$ referring to $\Lambda_c^+ \to \Lambda\pi^+\pi^0$ $(A')$ and $\Lambda_c^+ \to \Lambda K^+\pi^+\pi^-$ $(B)$ referring to $\Lambda_c^+ \to \Lambda\pi^+\pi^+\pi^-$ $(B')$, and signal yields for reference modes at different energy points. Uncertainties are statistical only.

| $\sqrt{s}$(GeV) | $10^2\,\frac{\varepsilon_A}{\varepsilon_{A'}}$ | $10^2\,\frac{\varepsilon_B}{\varepsilon_{B'}}$ | $N^{A'}$ | $N^{B'}$ |
|---|---|---|---|---|
| 4.599 | $72.3\pm0.5$ | $61.9\pm0.6$ | $1229.6\pm38.3$ | $517.3\pm24.8$ |
| 4.612 | $72.8\pm0.5$ | $61.0\pm0.6$ | $194.0\pm16.2$ | $98.5\pm10.7$ |
| 4.628 | $72.7\pm0.5$ | $61.4\pm0.6$ | $968.4\pm35.7$ | $398.2\pm22.4$ |
| 4.641 | $72.7\pm0.5$ | $62.0\pm0.6$ | $1117.5\pm38.6$ | $453.6\pm24.0$ |
| 4.661 | $72.9\pm0.5$ | $62.7\pm0.6$ | $952.4\pm35.7$ | $449.8\pm23.5$ |
| 4.682 | $73.5\pm0.5$ | $62.4\pm0.6$ | $3013.0\pm63.4$ | $1446.7\pm42.2$ |
| 4.699 | $73.9\pm0.5$ | $62.7\pm0.6$ | $846.9\pm33.8$ | $447.2\pm23.1$ |
| 4.740 | $73.8\pm0.5$ | $66.1\pm0.7$ | $312.2\pm20.3$ | $191.4\pm15.3$ |
| 4.750 | $74.7\pm0.6$ | $66.0\pm0.7$ | $595.9\pm28.5$ | $314.0\pm19.7$ |
| 4.781 | $75.8\pm0.6$ | $66.9\pm0.7$ | $839.4\pm33.7$ | $398.0\pm22.2$ |
| 4.843 | $76.6\pm0.6$ | $68.8\pm0.8$ | $587.0\pm28.8$ | $321.4\pm20.4$ |
| 4.918 | $77.8\pm0.7$ | $72.5\pm0.9$ | $262.9\pm18.8$ | $167.4\pm14.4$ |
| 4.950 | $80.2\pm0.7$ | $72.8\pm1.0$ | $166.1\pm15.2$ | $107.4\pm11.5$ |

The angle $\theta_{\rm truth}$ is defined as the opening angle between each reconstructed and the corresponding simulated tracks. The signal shape is derived from the events with $\theta_{\rm truth} < 20°$ for all tracks. For the signal decay modes, the parameters of the Gaussian functions are shared with those of the corresponding reference decay modes due to low number of events in the signal peaks.

For the signal decay modes, the background shapes consist of an ARGUS function [35] to describe the combinatorial components, a shape extracted from exclusive MC simulations to describe the remaining peaking background and a shape extracted from signal MC samples to describe the wrongly reconstructed events. The peaking backgrounds arise from the following specific decay processes: $\Lambda_c^+ \to \Xi^0 K^+$ for the $\Lambda_c^+ \to \Lambda K^+\pi^0$ channel, and $\Lambda_c^+ \to \Lambda K^+ K_S^0$ for the $\Lambda_c^+ \to \Lambda K^+\pi^+\pi^-$ channel. The corresponding yields of these peaking backgrounds are determined using exclusive MC samples. As there are no significant sources of peaking contamination for the reference modes, here the background is described with only an ARGUS function and the unmatched background shape. The ARGUS function has an endpoint fixed at $E_{\rm beam}$ and a floating slope parameter shared between signal modes and reference modes for better precision. Unmatched events, studied through the signal MC samples, shows a distribution that is not flat. In the simultaneous fit, the determination of yields associated with these unmatched events relies on evaluating the ratio between matched signal yields and unmatched background yields, where the ratio is determined from MC simulation.

By maximizing the likelihood of the simultaneous fit, we obtain

$$\mathcal{R}_{\Lambda K^+\pi^0} = \frac{\mathcal{B}(\Lambda_c^+ \to \Lambda K^+\pi^0)}{\mathcal{B}(\Lambda_c^+ \to \Lambda\pi^+\pi^0)} = (2.09\pm0.39)\times10^{-2}$$

and

$$\mathcal{R}_{\Lambda K^+\pi^+\pi^-} = \frac{\mathcal{B}(\Lambda_c^+ \to \Lambda K^+\pi^+\pi^-)}{\mathcal{B}(\Lambda_c^+ \to \Lambda\pi^+\pi^+\pi^-)} = (1.13\pm0.41)\times10^{-2},$$

where the uncertainties are statistical only.

## V. SYSTEMATIC UNCERTAINTY

The significant sources of systematic uncertainty in the $\mathcal{R}$ measurement comprise those associated with the tracking and PID of charged tracks, the $\pi^0$ reconstruction, the $\Delta E$ requirement, the simultaneous fit, MC modeling, the understanding of the peaking backgrounds and the performance of truth matching. The relative quantities associated with these uncertainties are detailed in Table II. The uncertainties from the total number of $\Lambda_c^+\bar{\Lambda}_c^-$ pairs and the BF of $\Lambda \to p\pi^-$ are canceled in the relative measurement.

We study the uncertainty from the tracking with a control sample of $e^+e^- \to K^+K^-\pi^+\pi^-$ events collected at $\sqrt{s} = 4.178$ GeV, where the tracking efficiency is measured both in data and MC simulation. We reweight the efficiency for kaon and pion according to their transverse momenta, and use the new efficiency to get the deviations in the obtained value of $\mathcal{R}$, which is 0.2% for $\Lambda_c^+ \to \Lambda K^+\pi^0$ and 1.2% for $\Lambda_c^+ \to \Lambda K^+\pi^+\pi^-$. Similarly, the uncertainty associated with PID is studied with control samples of $e^+e^- \to K^+K^-K^+K^-$, $K^+K^-\pi^+\pi^-$, $K^+K^-\pi^+\pi^-\pi^0$, $\pi^+\pi^-\pi^+\pi^-$, and $\pi^+\pi^-\pi^+\pi^-\pi^0$ events at $\sqrt{s} = 4.178$ GeV. We determine the PID efficiencies of kaon and pion identification both in data and MC simulations and reweight the efficiencies according to their momenta. In this way, the uncertainties for $\Lambda_c^+ \to \Lambda K^+\pi^0$ and $\Lambda_c^+ \to \Lambda K^+\pi^+\pi^-$ are 0.4% and 1.6%, respectively. The reconstruction efficiency of $\pi^0$ is studied through the $D \to K\pi\pi^0$ mode. Since the $\pi^0$ momentum in our signal and reference modes are not fully the same, we reweight them according to the $\pi^0$ momentum and obtain the associated uncertainty 0.8% for $\Lambda_c^+ \to \Lambda K^+\pi^0$.

In the nominal fit, the parameters of the Gaussian function are shared between the signal and reference modes, and the uncertainty associated with these parameters is neglected due to the clear signal in the reference modes. The uncertainty related to the background shape is assessed by varying the ARGUS end point by $\pm0.15$ MeV. The alternative fit results in an uncertainty of 1.2% for $\Lambda_c^+ \to \Lambda K^+\pi^0$ and 3.7% for $\Lambda_c^+ \to \Lambda K^+\pi^+\pi^-$.

The uncertainty due to the fixed contribution of the peaking background yields in the fit is investigated by varying the fixed yields within $\pm1\sigma$ of the PDG BFs of individual background sources. The largest differences observed in the fitted signal yield are assigned as the systematic uncertainties, which are 1.3% for $\Lambda_c^+ \to \Lambda K^+\pi^0$ and 2.6% for $\Lambda_c^+ \to \Lambda K^+\pi^+\pi^-$.

TABLE II. Systematic uncertainty in the relative BF measurements (in %).

| Source | $\mathcal{R}_{\Lambda K^+\pi^0}$ | $\mathcal{R}_{\Lambda K^+\pi^+\pi^-}$ |
|---|---|---|
| Tracking | 0.2 | 1.2 |
| PID | 0.4 | 1.6 |
| $\pi^0$ reconstruction | 0.7 | … |
| $M_{\rm BC}$ fit | 1.2 | 3.7 |
| $\Delta E$ requirement | 0.1 | 0.1 |
| MC model | 1.8 | 0.1 |
| Peaking background | 1.3 | 2.6 |
| Truth matching | 2.1 | 0.1 |
| Total | 3.4 | 4.9 |

The detection efficiency is determined after applying the $\Delta E$ requirements to the signal MC samples. Possible differences between the data and MC samples in the $\Delta E$ distributions are studied using the reference modes. The signal MC samples are smeared according to data, and the BF difference between nominal and smeared samples are assigned as the uncertainties, which are 0.1% for $\Lambda_c^+ \to \Lambda K^+\pi^0$ and 0.1% for $\Lambda_c^+ \to \Lambda K^+\pi^+\pi^-$.

The systematic uncertainties associated with MC modeling are evaluated by generating alternative signal MC samples. We add some possible resonances to the signal MC samples, for instance, $\Lambda_c^+ \to \Lambda K^*(892)^+$ and $\Lambda_c^+ \to \Sigma(1385)^0 K^*(892)^+$, in $\Lambda_c^+ \to \Lambda K^+\pi^0$ and $\Lambda_c^+ \to \Lambda K^+\pi^+\pi^-$, respectively. The efficiency differences obtained with the nominal and alternative signal MC samples are assigned as the uncertainties, which are 1.8% and 0.1% for $\Lambda_c^+ \to \Lambda K^+\pi^0$ and $\Lambda_c^+ \to \Lambda K^+\pi^+\pi^-$.

We investigate the uncertainty linked to the performance of the truth matching by varying the $\theta_{\rm truth}$ requirement by $20° \pm 1°$ in the signal MC. The relative differences obtained from these variations are then utilized to estimate the corresponding systematic uncertainties which are assigned to be 2.1% for $\Lambda_c^+ \to \Lambda K^+\pi^0$ decay and 0.1% for $\Lambda_c^+ \to \Lambda K^+\pi^+\pi^-$ decay.

The statistical significance of the signal is calculated by $S = \sqrt{-2\ln(\mathcal{L}_0/\mathcal{L}_{\rm max})}$, where $\mathcal{L}_{\rm max}$ and $\mathcal{L}_0$ are the maximal likelihood of the fits with and without the signal contribution, respectively. To account for additive systematic uncertainties, which include the $M_{\rm BC}$ fit, peaking background and performance of truth matching, and under the assumption of their independence, we obtain quadratic sums of 2.8% for $\mathcal{R}_{\Lambda K^+\pi^0}$ and 4.5% for $\mathcal{R}_{\Lambda K^+\pi^+\pi^-}$. Taking these systematic uncertainties into consideration, the signal significance is found to be $5.7\sigma$ for the $\Lambda_c^+ \to \Lambda K^+\pi^0$ decay and $3.1\sigma$ for the $\Lambda_c^+ \to \Lambda K^+\pi^+\pi^-$ decay.

## VI. SUMMARY

In this paper, we report the first observation of the SCS decay $\Lambda_c^+ \to \Lambda K^+\pi^0$ with a significance of $5.7\sigma$ and the first evidence of $\Lambda_c^+ \to \Lambda K^+\pi^+\pi^-$ with a significance of $3.1\sigma$. The BFs are measured relative to their CF counterparts, which are $\frac{\mathcal{B}(\Lambda_c^+\to\Lambda K^+\pi^0)}{\mathcal{B}(\Lambda_c^+\to\Lambda\pi^+\pi^0)} = (2.09 \pm 0.39_{\rm stat} \pm 0.07_{\rm syst}) \times 10^{-2}$ and $\frac{\mathcal{B}(\Lambda_c^+\to\Lambda K^+\pi^+\pi^-)}{\mathcal{B}(\Lambda_c^+\to\Lambda\pi^+\pi^+\pi^-)} = (1.13 \pm 0.41_{\rm stat} \pm 0.06_{\rm syst}) \times 10^{-2}$. By combining our measurements with the $\mathcal{B}(\Lambda_c^+ \to \Lambda\pi^+\pi^0)$ and $\mathcal{B}(\Lambda_c^+ \to \Lambda\pi^+\pi^+\pi^-)$ from the PDG [28], we obtain the BFs $\mathcal{B}(\Lambda_c^+ \to \Lambda K^+\pi^0) = (1.49 \pm 0.27_{\rm stat} \pm 0.05_{\rm syst} \pm 0.08_{\rm ref}) \times 10^{-3}$ and $\mathcal{B}(\Lambda_c^+ \to \Lambda K^+\pi^+\pi^-) = (4.13 \pm 1.48_{\rm stat} \pm 0.20_{\rm syst} \pm 0.33_{\rm ref}) \times 10^{-4}$. Two recent theoretical works which are based on the quark $SU(3)$ flavor symmetry predict the $\mathcal{B}(\Lambda_c^+ \to \Lambda K^+\pi^0)$ to be $(4.5 \pm 0.8) \times 10^{-3}$ [10] and $(3.5 \pm 0.6) \times 10^{-3}$ [11]. Our measured value deviates from these predictions by $3.5\sigma$ and $3.0\sigma$, respectively. Our result of $\mathcal{B}(\Lambda_c^+ \to \Lambda K^+\pi^+\pi^-)$ is consistent with the measurement by the *BABAR* experiment [13]. The precision of both $\mathcal{B}(\Lambda_c^+ \to \Lambda K^+\pi^0)$ and $\mathcal{B}(\Lambda_c^+ \to \Lambda K^+\pi^+\pi^-)$ measurements is currently dominated by the statistical uncertainty. Improved precision for these two SCS decays will be achievable from the larger datasets that are expected to be collected in the near future, following the upgrade of the BEPCII collider [19,36]. These improved measurements will provide valuable insights into the properties of these decays and help in refining our understanding of charmed baryon decays.

## ACKNOWLEDGMENTS

The BESIII Collaboration thanks the staff of BEPCII and the IHEP computing center for their strong support. This work is supported in part by National Key R&D Program of China under Contracts Nos. 2020YFA0406400, 2020YFA0406300; National Natural Science Foundation of China (NSFC) under Contracts No. 12205141, No. 11635010, No. 11735014, No. 11835012, No. 11935015, No. 11935016, No. 11935018, No. 11961141012, No. 12025502, No. 12035009, No. 12035013, No. 12061131003, No. 12192260, No. 12192261, No. 12192262, No. 12192263, No. 12192264, No. 12192265, No. 12221005, No. 12225509, and No. 12235017; the Chinese Academy of Sciences (CAS) Large-Scale Scientific Facility Program; the CAS Center for Excellence in Particle Physics (CCEPP); Joint Large-Scale Scientific Facility Funds of the NSFC and CAS under Contract No. U1832207; CAS Key Research Program of Frontier Sciences under Contracts No. QYZDJ-SSW-SLH003, No. QYZDJ-SSW-SLH040; 100 Talents Program of CAS; The Institute of Nuclear and Particle Physics (INPAC) and Shanghai Key Laboratory for Particle Physics and Cosmology; European Union's Horizon 2020 research and innovation programme under Marie Sklodowska-Curie grant agreement under Contract No. 894790; German Research Foundation DFG under Contracts No. 455635585, Collaborative Research Center CRC 1044, FOR5327, GRK 2149; Istituto Nazionale di Fisica Nucleare, Italy; Ministry of Development of Turkey under Contract No. DPT2006K-120470; National Research Foundation of Korea under Contract No. NRF-2022R1A2C1092335; National Science and Technology fund of Mongolia; National Science Research and Innovation Fund (NSRF) via the Program Management Unit for Human Resources & Institutional Development, Research and Innovation of Thailand under Contract No. B16F640076; Polish National Science Centre under Contract No. 2019/35/O/ST2/02907; The Swedish Research Council; U.S. Department of Energy under Contract No. DE-FG02-05ER41374.

M. Ablikim,[1] M. N. Achasov,[4,b] P. Adlarson,[75] X. C. Ai,[81] R. Aliberti,[35] A. Amoroso,[74a,74c] M. R. An,[39] Q. An,[71,58] Y. Bai,[57] O. Bakina,[36] I. Balossino,[29a] Y. Ban,[46,g] H.-R. Bao,[63] V. Batozskaya,[1,44] K. Begzsuren,[32] N. Berger,[35] M. Berlowski,[44] M. Bertani,[28a] D. Bettoni,[29a] F. Bianchi,[74a,74c] E. Bianco,[74a,74c] A. Bortone,[74a,74c] I. Boyko,[36] R. A. Briere,[5] A. Brueggemann,[68] H. Cai,[76] X. Cai,[1,58] A. Calcaterra,[28a] G. F. Cao,[1,63] N. Cao,[1,63] S. A. Cetin,[62a] J. F. Chang,[1,58] T. T. Chang,[77] W. L. Chang,[1,63] G. R. Che,[43] G. Chelkov,[36,a] C. Chen,[43] Chao Chen,[55] G. Chen,[1] H. S. Chen,[1,63] M. L. Chen,[1,58,63] S. J. Chen,[42] S. L. Chen,[45] S. M. Chen,[61] T. Chen,[1,63] X. R. Chen,[31,63] X. T. Chen,[1,63] Y. B. Chen,[1,58] Y. Q. Chen,[34] Z. J. Chen,[25,h] S. K. Choi,[10] X. Chu,[43] G. Cibinetto,[29a] S. C. Coen,[3] F. Cossio,[74c] J. J. Cui,[50] H. L. Dai,[1,58] J. P. Dai,[79] A. Dbeyssi,[18] R. E. de Boer,[3] D. Dedovich,[36] Z. Y. Deng,[1] A. Denig,[35] I. Denysenko,[36] M. Destefanis,[74a,74c] F. De Mori,[74a,74c] B. Ding,[66,1] X. X. Ding,[46,g] Y. Ding,[40] Y. Ding,[34] J. Dong,[1,58] L. Y. Dong,[1,63] M. Y. Dong,[1,58,63] X. Dong,[76] M. C. Du,[1] S. X. Du,[81] Z. H. Duan,[42] P. Egorov,[36,a] Y. H. Fan,[45] J. Fang,[1,58] S. S. Fang,[1,63] W. X. Fang,[1] Y. Fang,[1] Y. Q. Fang,[1,58] R. Farinelli,[29a] L. Fava,[74b,74c] F. Feldbauer,[3] G. Felici,[28a] C. Q. Feng,[71,58] J. H. Feng,[59] K. Fischer,[69] M. Fritsch,[3] C. D. Fu,[1] J. L. Fu,[63] Y. W. Fu,[1] H. Gao,[63] Y. N. Gao,[46,g] Yang Gao,[71,58] S. Garbolino,[74c] I. Garzia,[29a,29b] P. T. Ge,[76] Z. W. Ge,[42] C. Geng,[59] E. M. Gersabeck,[67] A. Gilman,[69] K. Goetzen,[13] L. Gong,[40] W. X. Gong,[1,58] W. Gradl,[35] S. Gramigna,[29a,29b] M. Greco,[74a,74c] M. H. Gu,[1,58] Y. T. Gu,[15] C. Y. Guan,[1,63] Z. L. Guan,[22] A. Q. Guo,[31,63] L. B. Guo,[41] M. J. Guo,[50] R. P. Guo,[49] Y. P. Guo,[12,f] A. Guskov,[36,a] J. Gutierrez,[27] T. T. Han,[1] W. Y. Han,[39] X. Q. Hao,[19] F. A. Harris,[65] K. K. He,[55] K. L. He,[1,63] F. H. H. Heinsius,[3] C. H. Heinz,[35] Y. K. Heng,[1,58,63] C. Herold,[60] T. Holtmann,[3] P. C. Hong,[12,f] G. Y. Hou,[1,63] X. T. Hou,[1,63] Y. R. Hou,[63] Z. L. Hou,[1] B. Y. Hu,[59] H. M. Hu,[1,63] J. F. Hu,[56,i] T. Hu,[1,58,63] Y. Hu,[1] G. S. Huang,[71,58] K. X. Huang,[59] L. Q. Huang,[31,63] X. T. Huang,[50] Y. P. Huang,[1] T. Hussain,[73] N. Hüsken,[27,35] N. in der Wiesche,[68] M. Irshad,[71,58] J. Jackson,[27] S. Jaeger,[3] S. Janchiv,[32] J. H. Jeong,[10] Q. Ji,[1] Q. P. Ji,[19] X. B. Ji,[1,63] X. L. Ji,[1,58] Y. Y. Ji,[50] X. Q. Jia,[50] Z. K. Jia,[71,58] H. J. Jiang,[76] P. C. Jiang,[46,g] S. S. Jiang,[39] T. J. Jiang,[16] X. S. Jiang,[1,58,63] Y. Jiang,[63] J. B. Jiao,[50] Z. Jiao,[23] S. Jin,[42] Y. Jin,[66] M. Q. Jing,[1,63] X. M. Jing,[63] T. Johansson,[75] X. K.,[1] S. Kabana,[33] N. Kalantar-Nayestanaki,[64] X. L. Kang,[9] X. S. Kang,[40] M. Kavatsyuk,[64] B. C. Ke,[81] V. Khachatryan,[27] A. Khoukaz,[68] R. Kiuchi,[1] R. Kliemt,[13] O. B. Kolcu,[62a] B. Kopf,[3] M. Kuessner,[3] A. Kupsc,[44,75] W. Kühn,[37] J. J. Lane,[67] P. Larin,[18] A. Lavania,[26] L. Lavezzi,[74a,74c] T. T. Lei,[71,58] Z. H. Lei,[71,58] H. Leithoff,[35] M. Lellmann,[35] T. Lenz,[35] C. Li,[47] C. Li,[43] C. H. Li,[39] Cheng Li,[71,58] D. M. Li,[81] F. Li,[1,58] G. Li,[1] H. Li,[71,58] H. B. Li,[1,63] H. J. Li,[19] H. N. Li,[56,i] Hui Li,[43] J. R. Li,[61] J. S. Li,[59] J. W. Li,[50] Ke Li,[1] L. J. Li,[1,63] L. K. Li,[1] Lei Li,[48] M. H. Li,[43] P. R. Li,[38,k] Q. X. Li,[50] S. X. Li,[12] T. Li,[50] W. D. Li,[1,63] W. G. Li,[1] X. H. Li,[71,58] X. L. Li,[50] Xiaoyu Li,[1,63] Y. G. Li,[46,g] Z. J. Li,[59] Z. X. Li,[15] C. Liang,[42] H. Liang,[1,63] H. Liang,[71,58] Y. F. Liang,[54] Y. T. Liang,[31,63] G. R. Liao,[14] L. Z. Liao,[50] Y. P. Liao,[1,63] J. Libby,[26] A. Limphirat,[60] D. X. Lin,[31,63] T. Lin,[1] B. J. Liu,[1] B. X. Liu,[76] C. Liu,[34] C. X. Liu,[1] F. H. Liu,[53] Fang Liu,[1] Feng Liu,[6] G. M. Liu,[56,i] H. Liu,[38,j,k] H. B. Liu,[15] H. M. Liu,[1,63] Huanhuan Liu,[1] Huihui Liu,[21] J. B. Liu,[71,58] J. Y. Liu,[1,63] K. Liu,[1] K. Y. Liu,[40] Ke Liu,[22] L. Liu,[71,58] L. C. Liu,[43] Lu Liu,[43] M. H. Liu,[12,f] P. L. Liu,[1] Q. Liu,[63] S. B. Liu,[71,58] T. Liu,[12,f] W. K. Liu,[43] W. M. Liu,[71,58] X. Liu,[38,j,k] Y. Liu,[81] Y. Liu,[38,j,k] Y. B. Liu,[43] Z. A. Liu,[1,58,63] Z. Q. Liu,[50] X. C. Lou,[1,58,63] F. X. Lu,[59] H. J. Lu,[23] J. G. Lu,[1,58] X. L. Lu,[1] Y. Lu,[7] Y. P. Lu,[1,58] Z. H. Lu,[1,63] C. L. Luo,[41] M. X. Luo,[80] T. Luo,[12,f] X. L. Luo,[1,58] X. R. Lyu,[63] Y. F. Lyu,[43] F. C. Ma,[40] H. Ma,[79] H. L. Ma,[1] J. L. Ma,[1,63] L. L. Ma,[50] M. M. Ma,[1,63] Q. M. Ma,[1] R. Q. Ma,[1,63] X. Y. Ma,[1,58] Y. Ma,[46,g] Y. M. Ma,[31] F. E. Maas,[18] M. Maggiora,[74a,74c] S. Malde,[69] Q. A. Malik,[73] A. Mangoni,[28b] Y. J. Mao,[46,g] Z. P. Mao,[1] S. Marcello,[74a,74c] Z. X. Meng,[66] J. G. Messchendorp,[13,64] G. Mezzadri,[29a] H. Miao,[1,63] T. J. Min,[42] R. E. Mitchell,[27] X. H. Mo,[1,58,63] B. Moses,[27] N. Yu. Muchnoi,[4,b] J. Muskalla,[35] Y. Nefedov,[36] F. Nerling,[18,d] I. B. Nikolaev,[4,b] Z. Ning,[1,58] S. Nisar,[11,l] Q. L. Niu,[38,j,k]

W. D. Niu,[55] Y. Niu,[50] S. L. Olsen,[63] Q. Ouyang,[1,58,63] S. Pacetti,[28b,28c] X. Pan,[55] Y. Pan,[57] A. Pathak,[34] P. Patteri,[28a] Y. P. Pei,[71,58] M. Pelizaeus,[3] H. P. Peng,[71,58] Y. Y. Peng,[38,j,k] K. Peters,[13,d] J. L. Ping,[41] R. G. Ping,[1,63] S. Plura,[35] V. Prasad,[33] F. Z. Qi,[1] H. Qi,[71,58] H. R. Qi,[61] M. Qi,[42] T. Y. Qi,[12,f] S. Qian,[1,58] W. B. Qian,[63] C. F. Qiao,[63] J. J. Qin,[72] L. Q. Qin,[14] X. S. Qin,[50] Z. H. Qin,[1,58] J. F. Qiu,[1] S. Q. Qu,[61] C. F. Redmer,[35] K. J. Ren,[39] A. Rivetti,[74c] M. Rolo,[74c] G. Rong,[1,63] Ch. Rosner,[18] S. N. Ruan,[43] N. Salone,[44] A. Sarantsev,[36,c] Y. Schelhaas,[35] K. Schoenning,[75] M. Scodeggio,[29a,29b] K. Y. Shan,[12,f] W. Shan,[24] X. Y. Shan,[71,58] J. F. Shangguan,[55] L. G. Shao,[1,63] M. Shao,[71,58] C. P. Shen,[12,f] H. F. Shen,[1,63] W. H. Shen,[63] X. Y. Shen,[1,63] B. A. Shi,[63] H. C. Shi,[71,58] J. L. Shi,[12] J. Y. Shi,[1] Q. Q. Shi,[55] R. S. Shi,[1,63] X. Shi,[1,58] J. J. Song,[19] T. Z. Song,[59] W. M. Song,[34,1] Y. J. Song,[12] Y. X. Song,[46,g] S. Sosio,[74a,74c] S. Spataro,[74a,74c] F. Stieler,[35] Y. J. Su,[63] G. B. Sun,[76] G. X. Sun,[1] H. Sun,[63] H. K. Sun,[1] J. F. Sun,[19] K. Sun,[61] L. Sun,[76] S. S. Sun,[1,63] T. Sun,[51,e] W. Y. Sun,[34] Y. Sun,[9] Y. J. Sun,[71,58] Y. Z. Sun,[1] Z. T. Sun,[50] Y. X. Tan,[71,58] C. J. Tang,[54] G. Y. Tang,[1] J. Tang,[59] Y. A. Tang,[76] L. Y. Tao,[72] Q. T. Tao,[25,h] M. Tat,[69] J. X. Teng,[71,58] V. Thoren,[75] W. H. Tian,[52] W. H. Tian,[59] Y. Tian,[31,63] Z. F. Tian,[76] I. Uman,[62b] Y. Wan,[55] S. J. Wang,[50] B. Wang,[1] B. L. Wang,[63] Bo Wang,[71,58] C. W. Wang,[42] D. Y. Wang,[46,g] F. Wang,[72] H. J. Wang,[38,j,k] J. P. Wang,[50] K. Wang,[1,58] L. L. Wang,[1] M. Wang,[50] Meng Wang,[1,63] N. Y. Wang,[63] S. Wang,[38,j,k] S. Wang,[12,f] T. Wang,[12,f] T. J. Wang,[43] W. Wang,[59] W. Wang,[72] W. P. Wang,[71,58] X. Wang,[46,g] X. F. Wang,[38,j,k] X. J. Wang,[39] X. L. Wang,[12,f] Y. Wang,[61] Y. D. Wang,[45] Y. F. Wang,[1,58,63] Y. L. Wang,[19] Y. N. Wang,[45] Y. Q. Wang,[1] Yaqian Wang,[17,1] Yi Wang,[61] Z. Wang,[1,58] Z. L. Wang,[72] Z. Y. Wang,[1,63] Ziyi Wang,[63] D. Wei,[70] D. H. Wei,[14] F. Weidner,[68] S. P. Wen,[1] C. W. Wenzel,[3] U. Wiedner,[3] G. Wilkinson,[69] M. Wolke,[75] L. Wollenberg,[3] C. Wu,[39] J. F. Wu,[1,8] L. H. Wu,[1] L. J. Wu,[1,63] X. Wu,[12,f] X. H. Wu,[34] Y. Wu,[71] Y. H. Wu,[55] Y. J. Wu,[31] Z. Wu,[1,58] L. Xia,[71,58] X. M. Xian,[39] T. Xiang,[46,g] D. Xiao,[38,j,k] G. Y. Xiao,[42] S. Y. Xiao,[1] Y. L. Xiao,[12,f] Z. J. Xiao,[41] C. Xie,[42] X. H. Xie,[46,g] Y. Xie,[50] Y. G. Xie,[1,58] Y. H. Xie,[6] Z. P. Xie,[71,58] T. Y. Xing,[1,63] C. F. Xu,[1,63] C. J. Xu,[59] G. F. Xu,[1] H. Y. Xu,[66] Q. J. Xu,[16] Q. N. Xu,[30] W. Xu,[1] W. L. Xu,[66] X. P. Xu,[55] Y. C. Xu,[78] Z. P. Xu,[42] Z. S. Xu,[63] F. Yan,[12,f] L. Yan,[12,f] W. B. Yan,[71,58] W. C. Yan,[81] X. Q. Yan,[1] H. J. Yang,[51,e] H. L. Yang,[34] H. X. Yang,[1] Tao Yang,[1] Y. Yang,[12,f] Y. F. Yang,[43] Y. X. Yang,[1,63] Yifan Yang,[1,63] Z. W. Yang,[38,j,k] Z. P. Yao,[50] M. Ye,[1,58] M. H. Ye,[8] J. H. Yin,[1] Z. Y. You,[59] B. X. Yu,[1,58,63] C. X. Yu,[43] G. Yu,[1,63] J. S. Yu,[25,h] T. Yu,[72] X. D. Yu,[46,g] C. Z. Yuan,[1,63] L. Yuan,[2] S. C. Yuan,[1] X. Q. Yuan,[1] Y. Yuan,[1,63] Z. Y. Yuan,[59] C. X. Yue,[39] A. A. Zafar,[73] F. R. Zeng,[50] S. H. Zeng,[72] X. Zeng,[12,f] Y. Zeng,[25,h] Y. J. Zeng,[1,63] X. Y. Zhai,[34] Y. C. Zhai,[50] Y. H. Zhan,[59] A. Q. Zhang,[1,63] B. L. Zhang,[1,63] B. X. Zhang,[1] D. H. Zhang,[43] G. Y. Zhang,[19] H. Zhang,[71] H. C. Zhang,[1,58,63] H. H. Zhang,[59] H. H. Zhang,[34] H. Q. Zhang,[1,58,63] H. Y. Zhang,[1,58] J. Zhang,[59] J. Zhang,[81] J. J. Zhang,[52] J. L. Zhang,[20] J. Q. Zhang,[41] J. W. Zhang,[1,58,63] J. X. Zhang,[38,j,k] J. Y. Zhang,[1] J. Z. Zhang,[1,63] Jianyu Zhang,[63] L. M. Zhang,[61] L. Q. Zhang,[59] Lei Zhang,[42] P. Zhang,[1,63] Q. Y. Zhang,[39,81] Shuihan Zhang,[1,63] Shulei Zhang,[25,h] X. D. Zhang,[45] X. M. Zhang,[1] X. Y. Zhang,[50] Y. Zhang,[69] Y. Zhang,[72] Y. T. Zhang,[81] Y. H. Zhang,[1,58] Yan Zhang,[71,58] Yao Zhang,[1] Z. D. Zhang,[1] Z. H. Zhang,[1] Z. L. Zhang,[34] Z. Y. Zhang,[43] Z. Y. Zhang,[76] G. Zhao,[1] J. Y. Zhao,[1,63] J. Z. Zhao,[1,58] Lei Zhao,[71,58] Ling Zhao,[1] M. G. Zhao,[43] R. P. Zhao,[63] S. J. Zhao,[81] Y. B. Zhao,[1,58] Y. X. Zhao,[31,63] Z. G. Zhao,[71,58] A. Zhemchugov,[36,a] B. Zheng,[72] J. P. Zheng,[1,58] W. J. Zheng,[1,63] Y. H. Zheng,[63] B. Zhong,[41] X. Zhong,[59] H. Zhou,[50] L. P. Zhou,[1,63] X. Zhou,[76] X. K. Zhou,[6] X. R. Zhou,[71,58] X. Y. Zhou,[39] Y. Z. Zhou,[12,f] J. Zhu,[43] K. Zhu,[1] K. J. Zhu,[1,58,63] L. Zhu,[34] L. X. Zhu,[63] S. H. Zhu,[70] S. Q. Zhu,[42] T. J. Zhu,[12,f] W. J. Zhu,[12,f] Y. C. Zhu,[71,58] Z. A. Zhu,[1,63] J. H. Zou,[1] and J. Zu[71,58]

(BESIII Collaboration)

[1]*Institute of High Energy Physics, Beijing 100049, People's Republic of China*
[2]*Beihang University, Beijing 100191, People's Republic of China*
[3]*Bochum Ruhr-University, D-44780 Bochum, Germany*
[4]*Budker Institute of Nuclear Physics SB RAS (BINP), Novosibirsk 630090, Russia*
[5]*Carnegie Mellon University, Pittsburgh, Pennsylvania 15213, USA*
[6]*Central China Normal University, Wuhan 430079, People's Republic of China*
[7]*Central South University, Changsha 410083, People's Republic of China*
[8]*China Center of Advanced Science and Technology, Beijing 100190, People's Republic of China*
[9]*China University of Geosciences, Wuhan 430074, People's Republic of China*
[10]*Chung-Ang University, Seoul 06974, Republic of Korea*
[11]*COMSATS University Islamabad, Lahore Campus, Defence Road, Off Raiwind Road, 54000 Lahore, Pakistan*
[12]*Fudan University, Shanghai 200433, People's Republic of China*
[13]*GSI Helmholtzcentre for Heavy Ion Research GmbH, D-64291 Darmstadt, Germany*

[14]Guangxi Normal University, Guilin 541004, People's Republic of China
[15]Guangxi University, Nanning 530004, People's Republic of China
[16]Hangzhou Normal University, Hangzhou 310036, People's Republic of China
[17]Hebei University, Baoding 071002, People's Republic of China
[18]Helmholtz Institute Mainz, Staudinger Weg 18, D-55099 Mainz, Germany
[19]Henan Normal University, Xinxiang 453007, People's Republic of China
[20]Henan University, Kaifeng 475004, People's Republic of China
[21]Henan University of Science and Technology, Luoyang 471003, People's Republic of China
[22]Henan University of Technology, Zhengzhou 450001, People's Republic of China
[23]Huangshan College, Huangshan 245000, People's Republic of China
[24]Hunan Normal University, Changsha 410081, People's Republic of China
[25]Hunan University, Changsha 410082, People's Republic of China
[26]Indian Institute of Technology Madras, Chennai 600036, India
[27]Indiana University, Bloomington, Indiana 47405, USA
[28a]INFN Laboratori Nazionali di Frascati, I-00044 Frascati, Italy
[28b]INFN Sezione di Perugia, I-06100 Perugia, Italy
[28c]University of Perugia, I-06100 Perugia, Italy
[29a]INFN Sezione di Ferrara, I-44122 Ferrara, Italy
[29b]University of Ferrara, I-44122 Ferrara, Italy
[30]Inner Mongolia University, Hohhot 010021, People's Republic of China
[31]Institute of Modern Physics, Lanzhou 730000, People's Republic of China
[32]Institute of Physics and Technology, Peace Avenue 54B, Ulaanbaatar 13330, Mongolia
[33]Instituto de Alta Investigación, Universidad de Tarapacá, Casilla 7D, Arica 1000000, Chile
[34]Jilin University, Changchun 130012, People's Republic of China
[35]Johannes Gutenberg University of Mainz, Johann-Joachim-Becher-Weg 45, D-55099 Mainz, Germany
[36]Joint Institute for Nuclear Research, 141980 Dubna, Moscow region, Russia
[37]Justus-Liebig-Universitaet Giessen, II. Physikalisches Institut,
Heinrich-Buff-Ring 16, D-35392 Giessen, Germany
[38]Lanzhou University, Lanzhou 730000, People's Republic of China
[39]Liaoning Normal University, Dalian 116029, People's Republic of China
[40]Liaoning University, Shenyang 110036, People's Republic of China
[41]Nanjing Normal University, Nanjing 210023, People's Republic of China
[42]Nanjing University, Nanjing 210093, People's Republic of China
[43]Nankai University, Tianjin 300071, People's Republic of China
[44]National Centre for Nuclear Research, Warsaw 02-093, Poland
[45]North China Electric Power University, Beijing 102206, People's Republic of China
[46]Peking University, Beijing 100871, People's Republic of China
[47]Qufu Normal University, Qufu 273165, People's Republic of China
[48]Renmin University of China, Beijing 100872, People's Republic of China
[49]Shandong Normal University, Jinan 250014, People's Republic of China
[50]Shandong University, Jinan 250100, People's Republic of China
[51]Shanghai Jiao Tong University, Shanghai 200240, People's Republic of China
[52]Shanxi Normal University, Linfen 041004, People's Republic of China
[53]Shanxi University, Taiyuan 030006, People's Republic of China
[54]Sichuan University, Chengdu 610064, People's Republic of China
[55]Soochow University, Suzhou 215006, People's Republic of China
[56]South China Normal University, Guangzhou 510006, People's Republic of China
[57]Southeast University, Nanjing 211100, People's Republic of China
[58]State Key Laboratory of Particle Detection and Electronics,
Beijing 100049, Hefei 230026, People's Republic of China
[59]Sun Yat-Sen University, Guangzhou 510275, People's Republic of China
[60]Suranaree University of Technology, University Avenue 111, Nakhon Ratchasima 30000, Thailand
[61]Tsinghua University, Beijing 100084, People's Republic of China
[62a]Turkish Accelerator Center Particle Factory Group, Istinye University, 34010 Istanbul, Turkey
[62b]Turkish Accelerator Center Particle Factory Group, Near East University,
Nicosia, North Cyprus, 99138, Mersin 10, Turkey
[63]University of Chinese Academy of Sciences, Beijing 100049, People's Republic of China
[64]University of Groningen, NL-9747 AA Groningen, The Netherlands
[65]University of Hawaii, Honolulu, Hawaii 96822, USA
[66]University of Jinan, Jinan 250022, People's Republic of China

[67]*University of Manchester, Oxford Road, Manchester M13 9PL, United Kingdom*
[68]*University of Muenster, Wilhelm-Klemm-Strasse 9, 48149 Muenster, Germany*
[69]*University of Oxford, Keble Road, Oxford OX13RH, United Kingdom*
[70]*University of Science and Technology Liaoning, Anshan 114051, People's Republic of China*
[71]*University of Science and Technology of China, Hefei 230026, People's Republic of China*
[72]*University of South China, Hengyang 421001, People's Republic of China*
[73]*University of the Punjab, Lahore-54590, Pakistan*
[74a]*University of Turin, I-10125 Turin, Italy*
[74b]*University of Eastern Piedmont, I-15121 Alessandria, Italy*
[74c]*INFN, I-10125 Turin, Italy*
[75]*Uppsala University, Box 516, SE-75120 Uppsala, Sweden*
[76]*Wuhan University, Wuhan 430072, People's Republic of China*
[77]*Xinyang Normal University, Xinyang 464000, People's Republic of China*
[78]*Yantai University, Yantai 264005, People's Republic of China*
[79]*Yunnan University, Kunming 650500, People's Republic of China*
[80]*Zhejiang University, Hangzhou 310027, People's Republic of China*
[81]*Zhengzhou University, Zhengzhou 450001, People's Republic of China*

[a]Also at the Moscow Institute of Physics and Technology, Moscow 141700, Russia.
[b]Also at the Novosibirsk State University, Novosibirsk 630090, Russia.
[c]Also at the NRC "Kurchatov Institute," PNPI, 188300 Gatchina, Russia.
[d]Also at Goethe University Frankfurt, 60323 Frankfurt am Main, Germany.
[e]Also at Key Laboratory for Particle Physics, Astrophysics and Cosmology, Ministry of Education; Shanghai Key Laboratory for Particle Physics and Cosmology; Institute of Nuclear and Particle Physics, Shanghai 200240, People's Republic of China.
[f]Also at Key Laboratory of Nuclear Physics and Ion-beam Application (MOE) and Institute of Modern Physics, Fudan University, Shanghai 200443, People's Republic of China.
[g]Also at State Key Laboratory of Nuclear Physics and Technology, Peking University, Beijing 100871, People's Republic of China.
[h]Also at School of Physics and Electronics, Hunan University, Changsha 410082, China.
[i]Also at Guangdong Provincial Key Laboratory of Nuclear Science, Institute of Quantum Matter, South China Normal University, Guangzhou 510006, China.
[j]Also at MOE Frontiers Science Center for Rare Isotopes, Lanzhou University, Lanzhou 730000, People's Republic of China.
[k]Also at Lanzhou Center for Theoretical Physics, Lanzhou University, Lanzhou 730000, People's Republic of China.
[l]Also at the Department of Mathematical Sciences, IBA, Karachi 75270, Pakistan.